\documentclass{emulateapj}
\usepackage{amsmath}
\usepackage{ulem} 
\newcommand{\pc}{{\rm pc}}

\newcommand{\Rl}{\rm r_0}
\newcommand{\jet}{\rm _j}
\newcommand{\disk}{\rm _d}
\renewcommand{\vec}[1]{\boldsymbol{#1}}
\newcommand{\uvec}[1]{\vec{\hat #1}}     
\newcommand{\cross}{\times}              
\newcommand{\itprod}{\! \cdot \!}        
\newcommand{\dop}{{\cal D}}              
\newcommand{\ilos}{_{\rm los}}
\newcommand{\los}{\uvec n_{\rm los}}     
\newcommand{\Rfl}{R_{\alpha}} 
\newcommand{\Rmap}{R_{\rm map}} 
\newcommand{\Iaol}{I_{\rm axis}/I_{\rm limb}}
\newcommand{\aelec}{\alpha_{\rm e}}   
\renewcommand{\plottwo}[2]{\plotone{#2}}

\begin{document}

\title{Synthetic synchrotron emission maps from MHD models for the jet of M87}


\author{J. Gracia}
\affil{Dublin Institute for Advanced Studies,
     31 Fitzwilliam Place, Dublin 2, Ireland}
\email{jgracia@cp.dias.ie}

\author{N. Vlahakis}
\affil{IASA and Section of Astrophysics, Astronomy and Mechanics,
     Department of Physics, University of Athens,
     Panepistimiopolis, GR--157\,84 Zografos, Athens, Greece}

\author{I. Agudo}
\affil{Instituto de Astrof\'isica de Andaluc\'ia (CSIC),
     Apartado 3004, E-18080 Granada, Spain}

\author{K. Tsinganos}
\affil{IASA and Section of Astrophysics, Astronomy and Mechanics,
     Department of Physics, University of Athens,
     Panepistimiopolis, GR--157\,84 Zografos, Athens, Greece}

\and

\author{S.~V. Bogovalov}
\affil{Moscow Engineering Physics Institute (State University),
     Kashirskoje shosse 31, 115409 Moscow, Russia}

\begin{abstract}
We present self-consistent global, steady-state MHD models and
synthetic optically thin synchrotron emission maps for the jet of
M87. The model consist of two distinct zones: an inner relativistic
outflow, which we identify with the observed jet, and an outer cold
disk-wind. While the former does not self-collimate efficiently due to
its high effective inertia, the latter fulfills all the conditions for
efficient collimation by the magneto-centrifugal mechanism. Given the
right balance between the effective inertia of the inner flow and the
collimation efficiency of the outer disk wind, the relativistic flow
is magnetically confined into a well collimated beam and matches the
measurements of the opening angle of M87 over several orders of
magnitude in spatial extent.  The synthetic synchrotron maps reproduce
the morphological structure of the jet of M87, i.e. center-bright
profiles near the core and limb-bright profiles away from the core. At
the same time, they also show a local increase of brightness at some
distance along the axis associated to a recollimation shock in the MHD
model. Its location coincides with the position of the optical knot
HST-1. In addition our best fitting model is consistent with a number
of observational constraints such as the magnetic field in the knot
HST-1, and the jet-to-counterjet brightness ratio.
\end{abstract}


\section{Introduction}
Since its discovery by \citet{Curtis} the jet of M87 is the classical
prototype for extragalactic jets. Due to its proximity at 16 Mpc
\citep{W+95, M+99, T+01} M87 is one of the 
closest radio galaxies, which allows present VLBI instruments to
resolve the transversal structure of the jet. Therefore, it is an
ideal candidate for testing specific jet formation mechanisms.  The
jet and its hot spots have been systematically studied across the
electromagnetic spectrum from the radio to X-rays, both, with
ground-based observations and from satellites \citep[for a review see
  e.g.][]{Biretta96}. The initial
opening angle is approximately 60\degr{} on scales of about 0.04 \pc{}
and decreases rapidly until reaching 10\degr{} at a
distance of 4 \pc{} from the core \citep{Biretta+02}. These
observations suggest that the jet of M87 is rather slowly collimated
across a length of several parsec.

The prevailing paradigm for jet formation and collimation is magnetic
self-collimation by the \citet{BP82} mechanism. This view is supported
by observations, which are consistent with the presence of a
non-vanishing toroidal magnetic field component \citep{A+02, A+08,
  Gabuzda+04, ZT05, G+A08}.  However, relativistic effects have been
shown to decrease the collimation efficiency, i.e. for a given
magnetic field configuration at the base of the jet (or rotator
efficiency), the fraction of total mass- and magnetic flux that is
asymptotically cylindrically collimated is lower for a relativistic
flow than for a non-relativistic flow \citep{BT99, TB00}. Not only is
the mass flux fraction lower, but the final opening angle is larger
for relativistic outflows due to the decollimating effect of the
electric field and the effective inertia of the plasma \citep{B01},
which both counter-act the pinching by the toroidal magnetic
field. However, this is strictly true only for initially radial
magnetic field structures, as opposed to extended
magnetic field configurations like disk-winds. See, e.g. 
\citet{FM01, VK04, KBVK07} for efficiently collimating MHD disk-wind models.

In a series of papers \citet{GTB05, TB05, TB02} suggested a steady-state
two-component MHD model. The model consists of an inner relativistic
outflow, which is identified with the observed jet, and an outer
non-relativistic disk-wind. While the inner relativistic jet is not
expected to collimate well through magnetic self-collimation, the very
same process operates efficiently in the outer
non-relativistic disk-wind. It is expected, that at least for a part
of the available parameter space, collimation in the outer disk-wind
is so efficient, that it might resist the decollimating inertia of the
inner relativistic plasma and channel the jet into a narrowly confined
beam. \citet{GTB05} have shown, that such two-component models could
easily account for the narrow appearance of the beam of extragalactic
jets by reproducing the observational measurement of the opening angle
distribution as a function of angular distance from the core by
\citet{Biretta+02}. Since in these models the jet, i.e. the
relativistic inner outflow, is strictly speaking not magnetically
self-collimated, but rather confined by the outer disk-wind, the
authors prefer to talk of collimation by magnetic confinement.

However, \citet{GTB05} could not fit the opening angle distribution with
a unique MHD model. Instead various sets of parameters reproduce the
observations with similar accuracy. 

In deriving the opening angle of their model, \citet{GTB05} made a
simplifying but crucial assumption. They identified the observed jet
with the inner relativistic outflow of their two-component model. More
specifically, the boundary of the jet was assumed to coincide with the
shape of a specific fieldline $\Psi_\alpha$, the one fieldline, which
separates the relativistic inner region from the non-relativistic
outer disk-wind at the base, or the launching surface, of the
outflow. So, the observed opening angle was strictly speaking fitted
by the shape of a single magnetic fieldline.

However, observations do not measure the plasma state directly, i.e.
in terms of velocity, temperature or magnetic field strength. Instead,
they register photon flux as a function of position on the plane of
the sky. As such, from an observational point of view, the width of
the jet is defined by the emission dropping below the detection limit
or a small fraction of the luminosity of the ridge line of the jet. So
the question is how well did \citet{GTB05} measure the width of
the jet in terms of observational quantities? Or more generally -- Do
MHD models explain the appearance of AGN jets?
The purpose of this paper is to answer exactly this question by
adopting the point of view of an observer. Assuming, that the main
radiation mechanism is synchrotron emission, we translate the steady-state MHD
model into a synthetic emission map and measure the width of the jet
using only these data. 

The outline of the paper is as follows.  In the following section
\ref{ch:MHD} we summarize the two-component MHD model for
extragalactic jets and discuss some of its properties relevant to this
work. In section \ref{ch:maps} we present compact expressions for the
calculation of synchrotron emission and apply these to our numerical
MHD models for the jet of M87.  Finally, we discuss our results and
draw some conclusions.

\section{MHD model}
\label{ch:MHD}
%
\begin{figure}
  \plottwo{f1_bw}{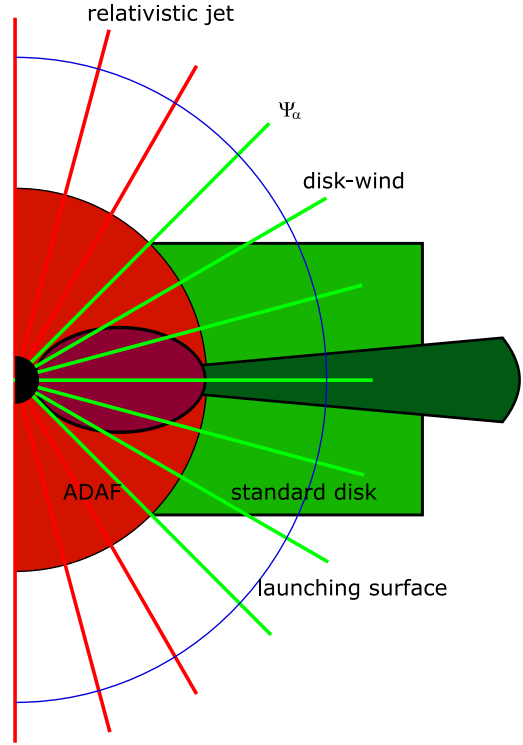}
  \caption{Illustrative sketch of the model (see text).}
  \label{fig:model}
\end{figure}

\begin{figure}
\plotone{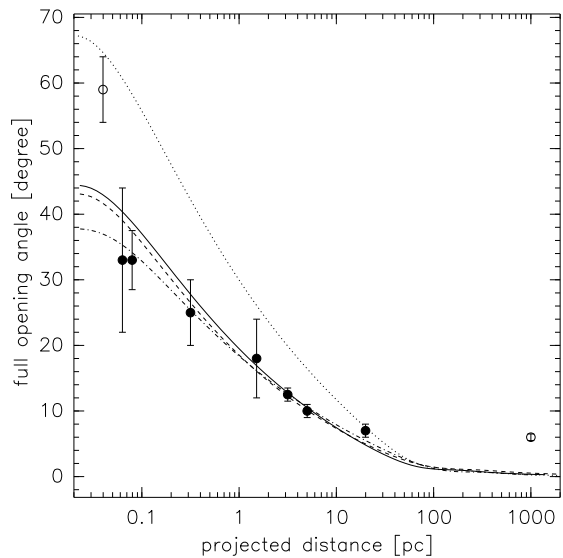}
  \caption{Comparison of the opening angle calculated from our MHD
    models and the observational data for M87. The lines show opening
    angle profiles as given by the separating fieldline $\Psi_\alpha$
    for model A ({\em solid line}), i.e. the best model, and model D
    ({\em dot-dashed}), which best reproduces the data based on
    $\Psi_\alpha$. The models B (({\em dashed})) and C ({\em dotted})
    are shown for completeness.  The data points marked by {\em filled
      circles} were taken into account in the fitting procedure. The
    innermost and outermost measurement ({\em open circles}) were
    disregarded. }
  \label{fig:opening_angle_fl}
\end{figure}

\begin{figure}
  \plottwo{f3_bw}{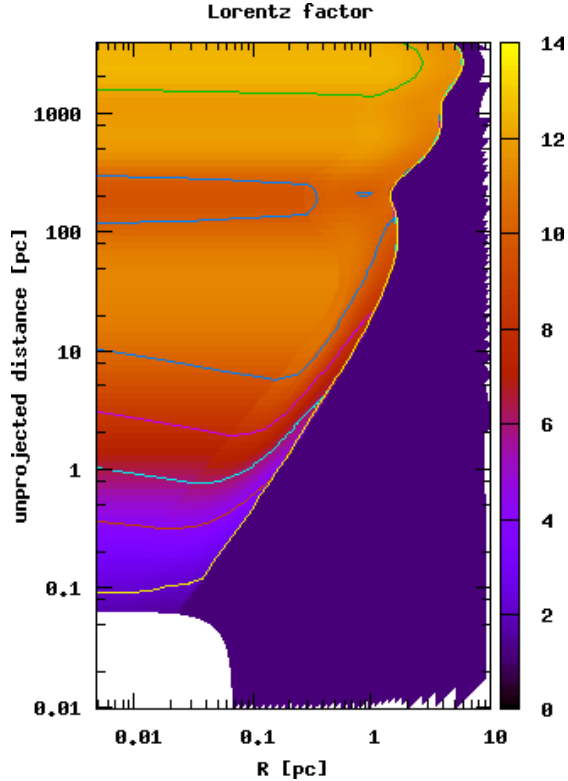}
  \caption{Lorentz factor along the flow for model A. Contour levels
    are shown at $\Gamma = 2,4,..,12$. The relativistic inner jet
    initially has Lorentz factors $\Gamma \sim 2-3$, but accelerates
    up to $\Gamma \sim 10-12$. Strong gradients may be seen across the
    jet and near the recollimation shock, if present.}
  \label{fig:acceleration}
\end{figure}

\begin{table*}
\begin{center}
  \caption{ \label{tab:params} Parameters and description of
    models.} \label{tab:models}
 \begin{tabular}{cccccccl}
    name & $\alpha\, [\degr]$ & $T\jet \, [mc^2]$ & $\Gamma\jet$ 
    & $\omega\jet \, [r_g/c]$ & $\Gamma\disk$ & $B_0$ & remarks\\
    \hline
    model A & 16.15 & 2.85 & 2.74 & 3.0 & 1.02 & 1.1 & best overall model\\
    model B & 14.45 & 3.0 & 2.74 & 3.0 & 1.02 & 0.9  & best fit to opening angle based on synchrotron map\\
    model C & 24.0 & 3.0 & 1.8 & 2.7 & 1.02 & 1.0     & highest limb-brightening and brightest knot HST-1\\
    model D & 14.4 & 3.45 & 3.26 & 3.0 & 1.01 & 1.27  & best fit to opening angle based on field line\\
    \tableline
    \multicolumn{8}{l}{The mass-flux rate of $10^{24} \,
      \mathrm{gs^{-1}}$ is a parameter common to all models.}
  \end{tabular}
\end{center}
\end{table*}

We adopt the model and notation of \citet{GTB05} and refer the reader
to that paper for details. A simple illustration of the model is shown
in Figure~\ref{fig:model}. It consist of two distinct zones; an inner
outflow, which is dominated by relativistic dynamics, and an outer
non-relativistic outflow. Both outflows originate from a spherical
launching surface located at a distance $\Rl$ from the black hole. The
launching surface is threaded by a helical magnetic field; the
poloidal component is initially perpendicular to the launching
surface.  The two zones are separated by a specific fieldline
$\Psi_\alpha$, where $2\alpha$ is the initial angular width, or
opening angle, of the inner relativistic outflow. In the following, we
will refer to these two distinct zones by relativistic jet and
(non-relativistic) disk-wind, respectively.

This two-component model is motivated by a similar two-component structure
of the underlying accretion flow consisting of an outer standard disk
\citep{SS73} and an inner hot plasma, which could be either an
advection dominated accretion flow \citep{NY94, PA97, GPKC03}, or the
final plunging region near the black hole, where relativistic dynamics
dominates through, e.g., frame-dragging, or the Blandford-Znajek process.

We impose two different sets of boundary conditions in the two
distinct zones along the launching surface. If the launching surface
is located beyond the fast magnetosonic surface, i.e. in the
hyperbolic MHD regime, the steady-state problem reduces to an initial
value Cauchy-type problem and the steady-state equations can be
integrated directly in terms of conserved integrals of motion as
described in \citet{TB02}. We stress, that beyond the launching
surface we solve the axisymmetric steady-state problem self-consistently, including the magnetic
field structure, as a function of the boundary values alone. However,
we do not solve the problem inside the launching surface, which is a
much more complicated exercise. A self-consistent solution of the full
problem needs to take into account the dynamics of the accretion flow,
something which is beyond the scope of this paper.

This procedure yields a set of quantities as a function of space in
the comoving frame of the jet.  The separating fieldline $\Psi_\alpha$
perfectly divides the relativistic from the non-relativistic
outflow. Inside of $\Psi_\alpha$ the plasma is highly relativistic,
both in terms of its bulk Lorentz factor, $\Gamma \gg 1$, and of its
thermal energy, $T \ge mc^2$, while outside the plasma is cold, $T \ll
mc^2$, and moving at non-relativistic speeds, $\Gamma = 1$. Also, the
magnetic field strength peaks close to the separating fieldline, where
fieldlines (and poloidal flowlines) of the inner outflow are strongly
compressed laterally and confined to a narrow sheet by the fieldlines
of the outer disk-wind thus forming a natural interface between both
zones.

\citet{GTB05} used the shape of the
separating fieldline $\Psi_\alpha$ to fit the observed opening angle
\citep{Biretta+02} as a function of distance from the core, however,
without taking into account projection effects, i.e. they assumed
implicitly, that the jet of M87 was in the plane of the sky.  There is
an on-going discussion on the inclination angle of the M87 jet
\citep[see e.g.][]{OHC89, RBJ+89, Biretta+99, LWJ07}. In this paper we
assume, that the jet of M87 is oriented at an angle
$\theta\ilos=40\degr$ from the line-of-sight as a compromise of values
discussed in the literature.

Unfortunately, the parameter space of the MHD model is degenerated in
the sense, that very different sets of parameters yield equally good
fits or even almost identical opening angle distributions.  Then, in
order to further constrain the model parameters we shall invoke the
radiation signatures of each model and compare them with the
corresponding observations. In this way, we may pin down a small
number of acceptable models which simultaneously satisfy the
constraints of the MHD model and also reproduce the observed
distribution of the emitted radiation.

We have run ~2600 axisymmetric steady-state MHD models and calculated synthetic synchrotron maps
for them. Here we discuss the four models that evaluate best under
different criteria. See table~\ref{tab:models} for their
parameters. Figure~\ref{fig:opening_angle_fl} compares the opening
angle as defined by the separating fieldline $\Psi_\alpha$, projected
under an angle of $\theta\ilos = 40\degr$, with the observational
measurements. Note that all our models fail to fit the jet width at
large distances close to the optical knot A at $\sim 900 \, \pc$ and
also at small distances close to the jet's origin, where there are
some uncertainties, as is discussed in the last section of the
paper. Models with large initial opening angle matching the first data
point as e.g. model C, have difficulties reproducing the opening angle
distribution. It is in general very difficult to make the curve more
concave. Models with initial opening angles falling slightly short of
the measured value (as models A, B, and D) may easily reproduce the
rest of the measurement and yield quite good fits. The quality of
agreement between the observed opening angle distribution and the
theoretical curve is measured by a simple $\chi^2$, i.e. the sum of
squares of the difference over the available observational data
points.  Model D best fits the opening angle at $\chi^2 = 2.9$, but
does not reproduce the morphological structure of the radiomaps
particularly well, as will be discussed in the next section. However,
the overall best model A reproduces the morphological structure and
still fits the opening angle data reasonably well with $\chi^2 =
7.8$. Note, that both models show a clear kink at $\sim 100 \,\pc$ due
to recollimation towards the axis.

Typically, models that fit the observed opening angle well are
moderately relativistic, both in terms of the initial outflow velocity
$\Gamma\jet \sim 2-3$ and the initial plasma internal energy $T\jet \sim 3
mc^2$.  In these models the plasma velocity increases along the flow
to values up to $\Gamma \sim 5-10$. However, the plasma may decelerate
and reaccelerate sharply at the recollimation shock. Strong
gradients of the plasma velocity may generally be present across the
jet, as seen in Figure.~\ref{fig:acceleration}. It is therefore very
difficult to assign a {\em typical} Lorentz factor to the whole jet.

\section{Synchrotron maps}
\label{ch:maps}

\begin{figure}
\plotone{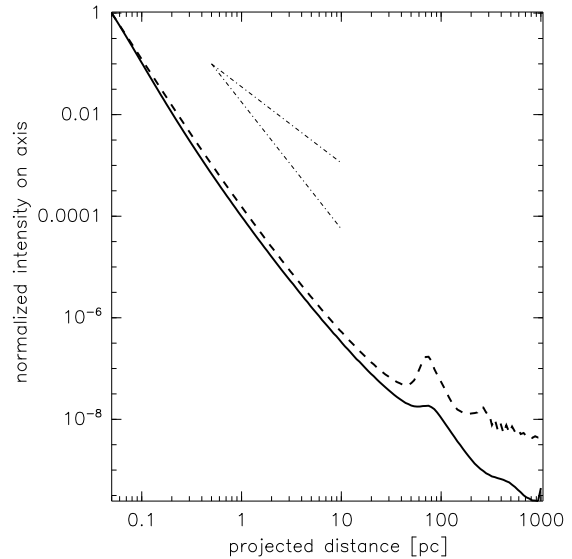}
  \caption{Comparison of the normalized synchrotron intensity along the jet axis
  calculated from model A ({\em solid lines}) and model C ({\em dashed
  lines}), respectively. The dashed lines indicate power-laws with
  index -1.5 and -2.5, respectively.} 
 \label{fig:trend}
\end{figure}

\begin{figure}
\plottwo{f5_bw}{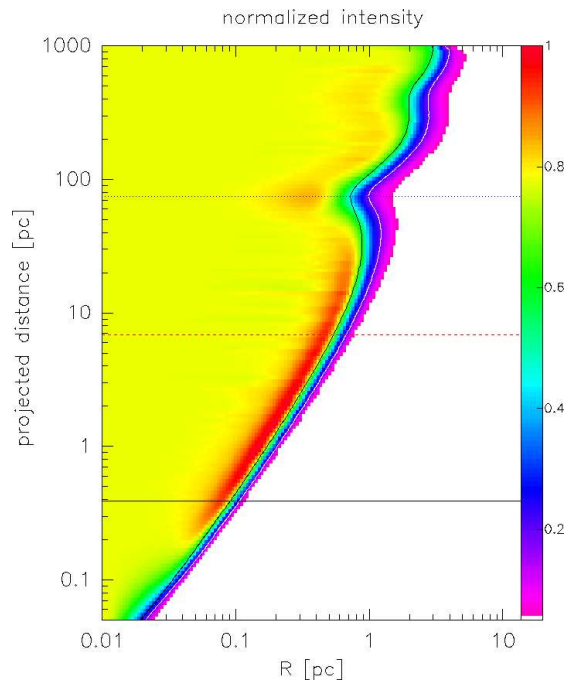}
\caption{Convolved synthetic synchrotron map. To increase contrast, 
the map has been divided by the trend along the jet axis
(Figure~\ref{fig:trend}). The two lines near the edge of the jet
indicate the jet width as defined by the separating fieldline ({\em
inner black line}) and the HWQM of the map ({\em outer white
line}). The three horizontal lines indicate the position of cuts
across the jet shown in Figure~\ref{fig:profile}.}
\label{fig:map}
\end{figure}

\begin{figure}
\plottwo{f6_bw}{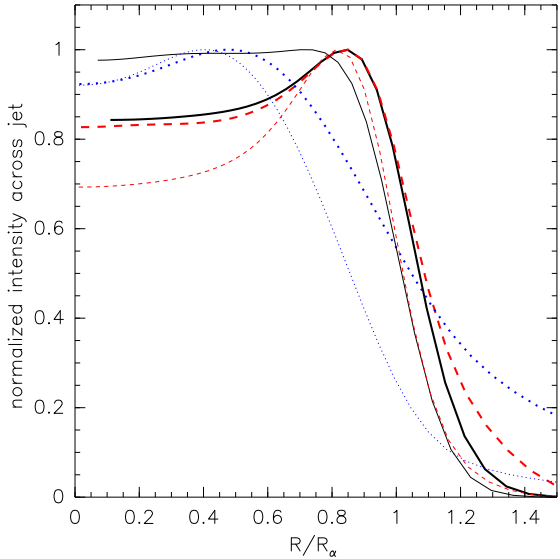}
\caption{Comparison of the normalized synchrotron intensity across the
  jet beam.  The overall best model A is indicated by heavy thick
  lines. For comparison model C, which has the most pronounced
  limb-brightening, is shown with thin lines. We plotted cuts across
  the jet beam at three different positions along the jet, i.e. close
  to the core ({\em solid/black line}), at intermediate distance ({\em
    dashed/red line}), and at large distance ({\em dotted/blue
    line}). See Figure~\ref{fig:map} for the location of the cuts. The
  lateral coordinate is normalized to the width of the jet $\Rfl$ as
  defined by the separating fieldline $\Psi_\alpha$ and the intensity
  is normalized to unity at the maximum.}
\label{fig:profile}
\end{figure}

\begin{figure}
\plotone{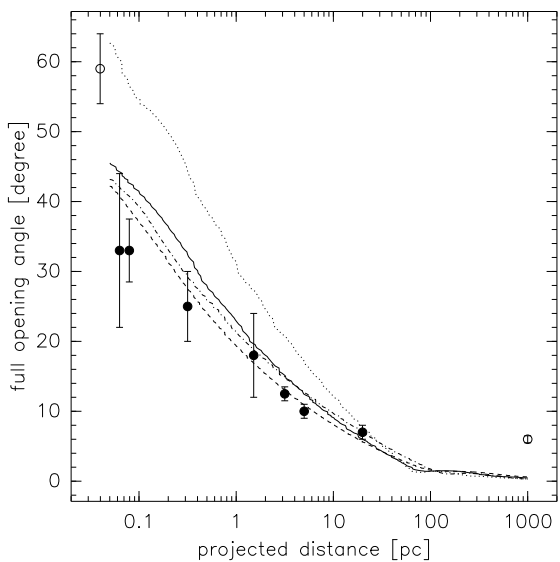}
  \caption{ Comparison of the opening angle calculated from synthetic
    maps and the observational data for M87. The lines show opening
    angle profiles as given by the HWQM contour $\Rmap$ of the overall
    best model A ({\em solid line}), and model B ({\em dashed line})
    which fits the opening angle data best.  Models C ({\em dotted})
    and D ({\em dot-dashed}) are shown for comparison.  Various
    symbols represent observational measurements.  The data points
    marked by {\em filled circles} were taken into account in the
    fitting procedure. The innermost and outermost measurement ({\em
      open circles}) were disregarded as explained in the text.}
  \label{fig:opening_angle_rm}
\end{figure}

\subsection{Calculation of synchrotron maps}

The calculation of the radio emissivity is done according the
relativistic generalization of expressions presented by \citet{L81}
and \citet{P70}. It is assumed that the radiating region is optically
thin with a uniform and isotropic distribution of electrons
\begin{equation}
  N'(E') \propto E'^{-(2\aelec+1)}
\end{equation}
giving rise to radiation with a spectral index $\aelec$=1, i.e.,
$S'_\nu \propto \nu'^{-\aelec}$. Note, that primed quantities are
measured in the comoving frame of the plasma, while unprimed
quantities refer to the lab frame or are independent of the frame of
reference.

We assume further, that the fraction of electron number density to
proton number density is constant throughout the emitting volume. Then
the number density of electrons $n_e = \int N(E) dE$ is proportional
to the density of the plasma, i.e. $n_e \propto \rho$ in every frame
of reference.

If the radiating plasma moves at relativistic speeds $\vec v$, i.e.
$\Gamma = 1/(1-\beta^2)^{1/2} \gg 1$ with $\beta=v/c$, it is
convenient to evaluate the synchrotron emission in the comoving frame
$\Sigma'$, instead of the lab- or observer frame $\Sigma$. Note
however, that the velocity of the comoving frame is not constant
within the emitting region, unless the velocity field $\vec v$ is
homogeneous, which is not true in our model. The synchrotron
emissivity in the comoving frame $\epsilon'$ is related to the
magnitude of the magnetic field perpendicular to the line-of-sight and
the electron density as
\begin{equation} \label{eq:em_c}
  \epsilon' \propto \rho' |\vec B' \cross  \los'|^{\aelec+1},
\end{equation}
where $\rho'$,$\vec B'$ and $\los'$ are the electron density, magnetic
field and line-of-sight unit vector in the comoving. These are given
in terms of the observers frame quantities through the
Lorentz-transformations as,
\begin{equation} \label{eq:denLT}
\rho' =  \rho/\Gamma, 
\end{equation}
\begin{equation} \label{eq:BLT}
\vec B' = \frac 1{\Gamma} \vec B 
    + \frac{\Gamma}{\Gamma+1} \frac{\vec v}{c^2}(\vec v \itprod \vec B),
    \qquad \vec E' = 0
\end{equation}
\begin{equation} \label{eq:nLT}
  \los' = \dop \los - (\dop+1)\frac{\Gamma}{\Gamma+1}
  \frac{\vec v}{c}. 
\end{equation}
We have introduced the Doppler factor ${\cal D} = (\Gamma(1-\vec v
\itprod \los/c))^{-1}$ for a compact notation and exploited the fact,
that in ideal MHD Ohm's law holds as $\vec E = -\vec v \cross \vec
B$ in both frames of reference.

The emissivity in the lab frame $\epsilon$ appears Doppler-boosted as
\begin{equation} \label{eq:em}
  \epsilon = {\cal D}^{\aelec+2} \, \epsilon'. 
\end{equation}
The amount of relativistic beaming strongly depends on the
line-of-sight angle, i.e. the angle between the jet axis and the
direction to the observer, which we fix to $\theta\ilos = 40\degr$.
 Finally, the flux in the
plane of the sky, $I$, is given by integration along the line-of-sight
\begin{equation} \label{eq:I}
  I = \int \epsilon \, d\ell\ilos.
\end{equation}

The synthetic synchrotron maps are convolved with a Gaussian beam to
qualitatively match the finite resolution of observed maps. However,
since our synthetic maps span more than four orders of magnitude in
distance from the core, the width of the Gaussian beam is not kept
constant. Typical radiomaps have a spatial resolution corresponding to
a couple of observing beams across the width of the jet. We therefore
use at each distance along the jet a Gaussian convolution kernel of
1-sigma width equal to a tenth of the jet radius at that distance,
i.e. $\sigma(Z) = \Rfl(Z)/10$.

The measurements of the opening angle collected by \citet{Biretta+02}
typically define the jet radius at distance $Z$ as the {\em half-width
at quarter-maximum} (HWQM) for the emission across the jet. We adopt
this definition and refer to it as the jet width of the radiomap
$2\Rmap$, with
\begin{equation}
\label{eq:Rmap}	
	I(\Rmap, Z) = max(I(R, Z))/4.
\end{equation}

\subsection{The best model}

The synthetic synchrotron emission maps shall qualitatively reproduce
two observational constraints: (i) the opening angle defined
through HWQM (half-width at quarter-maximum) in terms of small
$\chi^2$ values, and (ii) the pronounced limb-brightening in terms of
small values for the mean intensity on the axis over intensity on the
limb, i.e. $\left< \Iaol\right>$. As a secondary criterion, we favor
models showing some degree of enhanced emission close to the nominal
position of HST-1 at 70 pc.

We calculated synthetic synchrotron emission maps and evaluated them
according to our two criteria. For each of the two criteria we
assigned a score between 0..1 from a sorted list of values for
$\chi^2$ and $\left< \Iaol\right>$, respectively, and added those to
obtain the total score. A few formally high scoring models were
discarded because they did not show clear sign of increased brightness
at distances 70-200 pc, that could be identified with the knot
HST-1. The first model satisfying all criteria will be referred to as
{\em the best model} or simply {\em model A}.

The jet in the best model A is launched with a bulk velocity of
$\Gamma\jet=2.74$ and temperature $T=2.85 \,mc^2$. The separating
field line threads the launching surface radially at an angle $\alpha
= 16\degr$ with an angular velocity $\omega\jet = 3 \,r_g/c$. The
outer cold disk-wind is launched with velocity $\Gamma\disk = 1.02$.

For the best model A we plotted spatial 2-dimensional synthetic
synchrotron emission maps (Figure~\ref{fig:map}), the intensity measured
on the jet axis as a function of distance from the core
(Figure~\ref{fig:trend}), and the intensity profiles across the jet at
various positions along the jet axis (Figure~\ref{fig:profile}).

In Figure~\ref{fig:trend} we plot the trend along the axis for the
best model A and compare it with model C, which shows the highest
local brightness enhancement at the position of HST-1 in our
sample. The intensity along the axis is well described by a power-law
in projected distance. Both models have similar power-law indices
$\sim -2.5$. The model jets dim out faster than the observed jet whose
power-law index can be estimated to $\sim -1.5$ from the contrast of
several published intensity maps. However, this discrepancy is not
surprising as our model does not take into account any micro-physics
in order to re-energize the electron distribution. Both models show a
rise of luminosity around 70 \pc{} over the local power-law by a
factor of $\sim 3$ and $\sim 10$, respectively. The location of this
bright spot coincides with the location of strong recollimation
shocks, where density and magnetic field strength increase.

Figure~\ref{fig:map} shows the synthetic synchrotron map. In order to
increase the contrast we divided the map by the trend along the jet
axis, i.e. $I(R,Z)/I(0, Z)$. The jet beam is well defined, showing
large opening angles close to the core and becoming almost conical at
large distance. At the position of the recollimation shock the jet
width decreases and forms a visible neck.

The synthetic synchrotron maps show strong gradients of intensity
across the jet. These are more clearly visible in
Figure~\ref{fig:profile}, were cuts across the jet are shown. For
comparison we show also profiles for model C which has the most
pronounced limb-brightening.

The best model A shows limb-brightening already at a distance $\sim
0.2 \, \pc$ from the core. In the limb-bright region
$\left<\Iaol\right> = 0.76$ on average before convolution. After
convolution, the ratio rises to typically $0.85$. Model C bifurcates
at $\sim 0.5 \, {\rm pc}$ and has $\left<\Iaol\right> = 0.67$ and
$0.80$ before and after convolution, respectively.  These values agree
with those from observations by \citet{LWJ07} $\sim 0.63$,
\citet{K08Alaska} $\sim 0.6$, and \citet[and private
  communication]{K+07} $0.6 - 0.8$, depending on jet region and
resolution.  We note however, that the limb-brightness, as well as the
bifurcation distance depend on the details of convolution as discussed
later on.

Figure~\ref{fig:opening_angle_rm} compares the opening angle derived
from the synthetic maps $\Rmap$ with the observations. Model B fits
the observations best with $\chi^2=5.7$, but fails to reproduce the
limb-brightening. The best model A has $\chi^2 = 14.6$, which we
consider acceptable, in particular, if one considers the figure.

Finally, we compare the width of the jet from the MHD models, $\Rfl$,
with the jet width of the synthetic synchrotron maps, $\Rmap$. Both
curves are superimposed on the synthetic map in Figure~\ref{fig:map}.
For many models in our sample these two curves are virtually
indistinguishable. However, for our overall best model, these two
curves are noticeably different, even if they run almost
parallel. For this particular model $\Rfl$ underestimates the opening
angle by $\sim 20\%$ for all the length of the jet, i.e. $\Rmap \sim 1.2
\Rfl$. For other models in our sample, the difference is typically not more 
than $30\%-40\%$.

\subsection{The origin of limb-brightening}

\begin{figure}
\plottwo{f8_bw}{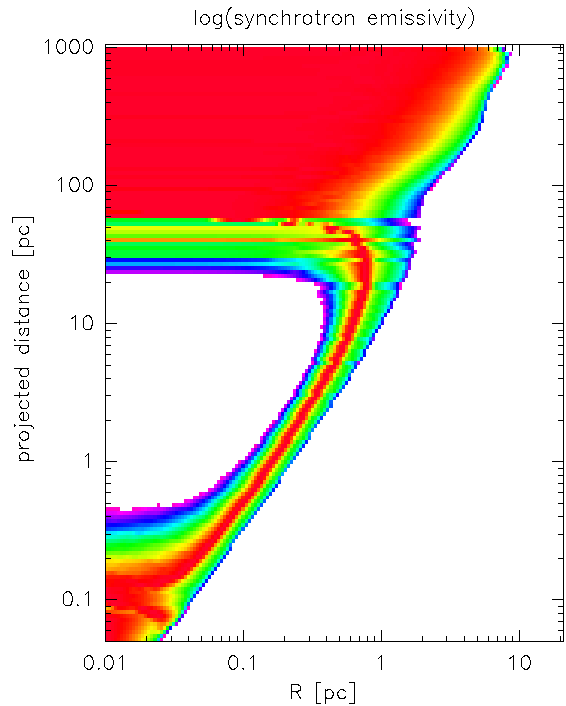}
  \caption{Synchrotron emissivity in the mid plane of the jet.  To
    increase contrast we normalized each row at distance $Z$ by the
    maximum emissivity across the jet at that particular distance.
    Close to the core the whole jet body contributes to the total
    emission along the line-of-sight. Further down the flow, most of
    the synchrotron emission originates from a thin shell at the outer
    edge of the jet, causing the bifurcation of the jet from
    center-bright to limb-bright.} \label{fig:emission_estimate}
\end{figure}

The synthetic emission map of the best model A in Figure~\ref{fig:map}
shows clear limb-brightening from a distance of $\sim 0.5 \, \pc$ up
to the location of the recollimation shock at $\sim 70 \, \pc$ and to
a lesser extend even beyond. This can be seen more clearly in
Figure~\ref{fig:emission_estimate}, which shows the synchrotron
emissivity $\epsilon'$ in the mid plane of the jet.  It is worth
noting, that close to the origin the whole jet body emits synchrotron
radiation. Further down the jet -- and certainly beyond 0.2 pc
projected distance from the core -- the emission is dominated by a
thin shell at the outer edge of the jet, while the inner region close
to the axis remains relative dark by a factor of more than a
hundred. Within our model this is easily explained noting that the
synchrotron emissivity equation~(\ref{eq:em_c}) is roughly
proportional to $\rho' \, B'^2$. Fieldlines within the relativistic
jet start out radially from the launching surface, but are soon
deflected by the outer cold disk wind. This leads to strong
concentration of magnetic flux, and therefore high field strength in a
narrow region at the interface between both components of the MHD
model. The simultaneously occurring increase of density plays only a
minor role.

Further down the flow, at around 60 pc, the emissivity within the jet
body increases again significantly and edge-brightening is less
pronounced. This region coincides well with the location of the
recollimation shock, where the topology of the magnetic field and the
distribution of plasma across the jet change significantly
again. Beyond the recollimation shock, the emissivity is almost
homogeneous across the jet.

However, the full picture is more complicated than that, since the
orientation of the highly non-homogeneous velocity field relative to
the line-of-sight is at least equally important. Not only does the
poloidal velocity vary from one magnetic flux surface to the next, but
the jet plasma is also rotating. Together this makes the Doppler
factor and the angle to the line-of-sight in the comoving frame highly
variable even along the circumsphere of the emitting plasma
shell. Cross sections perpendicular to the apparent axis of the jet
reveal, that the emissivity along the line-of-sight may be highly
asymmetric with respect to fore- and background halves of the jet.

\section{Discussion}

\begin{figure*}
\plottwo{f9_bw}{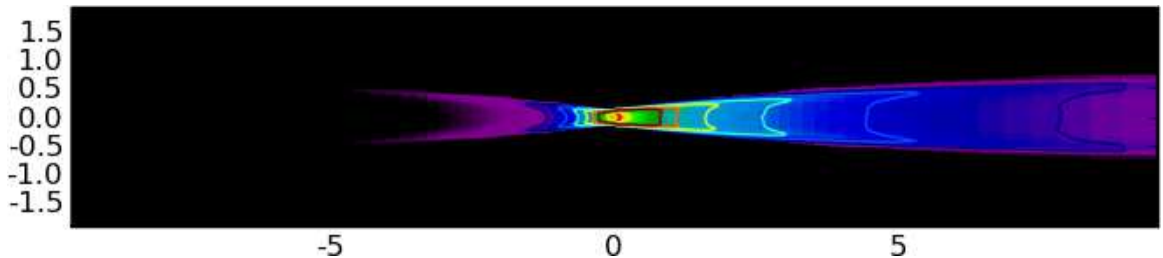}
  \caption{Synthetic synchrotron map for the inner jet of M87
    including the counterjet convolved with a constant Gaussian beam
    of $\sigma=0.03 \mathrm{pc}$. Unlike Fig~\ref{fig:map} this map has
    not been divided by the trend along the axis, but shows the full
    dynamic intensity contrast of $50000$ on a logarithmic scale. The
    contour lines are chosen as indicated on the colorbar in order to
    highlight the limb-brightening. The counterjet is under-luminous
    by factor of $\sim 13$. This map is meant to be compared to Fig~1
    in \citet{K+07}.} \label{fig:map_obs}
  \end{figure*}

Limb-brightening strongly constrains the parameters of MHD models for
the jet of M87. It is not only difficult to find parameters resulting
in a pronounced limb-brightening in terms of low $\left<\Iaol\right>$,
but also reproducing the spatial extend of the limb-bright
region. Depending on frequency and resolution, radiomaps for M87 show
limb-brightening from large distances almost right down to the core
\citep{CHS07, K+07}. However, the bifurcation distance, i.e. the
location where the jet morphology transits from center-bright to
limb-bright, seems to depend strongly on the frequency of a particular
observation. It is not clear {\em a priory} to what extent this is due
to intrinsic different emission properties at different observing
frequencies, or due to spatial resolution effects.  Y.~Y.~Kovalev
(private communication) has performed an analysis of the deep 15~GHz
VLBA image of the inner jet in M87 \citep{K+07}.  The
distance from the core at which the bifurcation becomes detectable
changes from 5~mas at the original 15~GHz resolution to more than
100~mas for a tenfold larger beam corresponding to 1.5~GHz. We
conclude, that variable resolution can in principle account for
different bifurcation distances at different frequencies.

Limb-brightening is often explained by stratified jets consisting of a
fast spine and a slow sheath \citep[see e.g.][]{AGI++00, GTC05}. In
those models the slow sheath brightens up in comparison to the spine
because of Doppler boosting into the $1/\Gamma$ cone. For the given
large line-of-sight angle of the M87 jet, the fast spine is seen from
well outside of the Doppler-boosted cone and its emission deboosted.
Here we suggest an alternative explanation for limb-brightening in
AGN, in particular M87. In our model the limb brightens up due to the
concentration of magnetic flux at the interface between the
relativistic jet and the confining non-relativistic disk-wind. The
synchrotron emissivity is therefore already intrinsically higher in
the comoving frame and limb-brightening does not rely on relativistic
aberration alone. Incidentally, as seen in Figure~\ref{fig:acceleration}
in our models the spine may at times be moving slower than plasma
further out which will counteract limb-brightening for large
inclination angles.

The exact value of the line-of-sight angle for the jet of M87 is still
under debate.  In this study we have assumed an angle
$\theta\ilos = 40\degr$ which brings it into the upper end of the
suggested range of values. In order to study the effect of the
line-of-sight angle on limb-brightening we have calculated synthetic
maps for model A under an angle $\theta\ilos=30\degr$. In comparison to the
original map, the bifurcation distance moves out to $0.5 \, \pc$,
limb-brightening is more pronounced, and the recollimation shock is
brighter with regard to the trend along the axis. At the same time
this obviously enlarges the apparent opening angle along the jet and
brings the recollimation shock closer to the core in disagreement with
observations. The simplest way to bring the opening angle curve back
down and move the recollimation shock out, is to decrease the model's
initial opening angle $\alpha$. However, models with lower $\alpha$
tend to show weaker limb-brightening and dimmer recollimation shocks
(if present at all). It may therefore be challenging to construct
models that reproduce all criteria for smaller inclination angles.

Under the assumption that jet and counterjet are intrinsically
  identical, one can in principle constrain the inclination angle by
  measuring the jet-to-counterjet brightness ratio $I_+/I_-$ given by
\begin{equation} \label{eq:counterjet}
  I_+/I = \left( 
  \frac{1 + \beta \, \cos \theta\ilos}{1 - \beta \, \cos \theta\ilos}
    \right)^{\aelec+2}, 
\end{equation}
where $\beta = v/c$ is the velocity in units of the speed of light and
$\aelec$ the electron distribution power-law index.  \citet{K+07}
estimated this ratio to be of the order 10 - 15, while \citet{LWJ07}
measured $I_+/I_- = 14.4$ and constrained the bulk flow velocity to
$\beta \sim 0.6 - 0.7 $. We calculated a synthetic synchrotron map for
the best model A pointing in the opposite direction, i.e. the
counterjet at $\theta\ilos = 220 \degr$, and estimated $I_+/I_- \sim
13$ from the on-axis intensities close to the core on either side of
it (see Fig~\ref{fig:map_obs}). The agreement is remarkable, in
particular taking into account, that one would naively expect a much
higher value ($> 200$) from the model's initial Lorentz factor
$\Gamma\jet = 2,74$, i.e. $\beta\jet = 0.93$, given in
table~\ref{tab:params}. However, eq.~(\ref{eq:counterjet}) assumes a
homogeneous velocity field along the axis and identical intrinsic
emissivities on either side of the core. Stratification of the jet
beam and rotation of the plasma about the axis, as is the case in our
models, can therefore lead to serious misinterpretation if not taken
into account.

While fitting the opening angle to the observational measurements
collected by \citet{Biretta+02}, we disregarded the outermost data
point close to the optical knot A at $900 \, \pc$. All our models fail
to reproduce this measurement and disregarding it made the
least-squares-fit more reliable and robust for the remaining data
points. However, we argue, that at such large distances the
interaction between the environment and the jet, which cannot be taken
into account by our numerical MHD solver, begins to dominate the jet
dynamics and thus renders our model insufficient. We have also
disregarded the innermost data point of $60 \degr$. This old
measurement has been challenged recently. \citet{LWJ07} state that the
opening angle at the jet base is larger than $15 \, \degr$ and
therefore consistent with \citet{Biretta+02} old measurement, but than
interestingly continue to repeat arguments that it might be an
overestimation. Maps presented by \citet{K+08} and \citet{WLJH08} seem
to leave room for smaller opening angles at the base. In general, the
initial opening angle measured from our synthetic emission maps can
easily be higher than $40\degr$.

Most of our models that reproduce the measured opening angle as a
function of distance as well as the limb-brightening, do at the same
time exhibit a recollimation shock\footnote{We also have a small range
  of models, that reproduces the opening angle without showing a
  strong recollimation shock (see \citealp{GTB05}, Figure~2), but
  those models do in general not show clear limb-brightening.} showing
some degree of locally enhanced emission. We identify this
recollimation shock with the bright knot HST-1. In most of our models
the emission at the bright knot does not increase by more than a
factor of 10 compared to the general trend, while for the real
\mbox{HST-1} the increase in emission level is probably
higher. However, this is not surprising, since our MHD model does not
include any micro-physics as e.g. particle acceleration, that could be
responsible for further enhancing the emission level. It is still
noteworthy, that the conditions in the vicinity of the recollimation
shock of the best model A, are similar to those postulated from SSC
models in order to fit the TeV emission in M87. In particular
\citet{H+03} recently presented a SSC model assuming, that the TeV
emission was originating from HST-1. In our model the Doppler factor
in the vicinity of the recollimation shock is rather low ${\cal D}
\sim 1 - 2$. For those Doppler factors, \citet{H+03} calculated a
magnetic field strength of $B' = (13 - 3.7) \; {\rm mG}$, which
matches very well $B'\sim 10 \; {\rm mG}$ in our model.

\section{Summary and conclusions}

We have calculated self-consistent global MHD models and synthetic
optically thin synchrotron maps with the aim to reproduce the reported
opening angle distribution as well as the morphological structure, in
particular limb-brightening, of the jet of M87. We have applied
different criteria to quantify the agreement between models and
observations. All criteria can be satisfied to high degree by
individual models in our database. However, no model does satisfy 
simultaneously all criteria exceedingly well. We identified a {\em best
  model} as compromise between the opening angle distribution and the
high degree of limb-brightening. This model satisfactorily meets our
criteria, but does not perform best in any single one.

\citet{GTB05} used the shape of the fieldline separating the two regions
on the boundary surface to define the opening angle. As shown in this
work, this particular fieldline and flux contours of synthetic maps
are parallel for a wide range of models and distances along the
jet. However, the first method may under-estimate the opening angle
for some parameters by typically $30\%$.

The morphological structure across the jet can be reproduced in
principle. A wide range of models show limb-brightening away from the
core. The transition from center-bright to limb-bright, the
bifurcation of the jet, generally occurs well below $1 \, {\rm pc}$
projected distance. A quantitative comparison with specific
observations, however, requires careful modeling of the convolving
beam and may require including opacity effects, i.e. optically thick
emission. Both effects are frequency dependent.

In our models limb-brightening is due to a strong increase of the
comoving frame synchrotron emissivity in a relatively thin shell near
the outer edge of the visible jet as result of a concentration of
magnetic flux at the interface between the relativistic outflow and
the non-relativistic disk-wind, i.e. the environment. In contrast to
other models limb-brightening here is not an immediate result of
deboosted emission in a slow moving sheath; in fact the 'sheath' in
our models typically moves slightly faster than the jet body.

The optical knot HST-1 has received much attention. Most MHD models
with acceptable fits to the opening angle distribution feature a shock
at the distance 70-200 pc. This is particularly true for models with
large initial opening angle $\alpha$. The shock is due to the fact,
that the jet is not in lateral force equilibrium at the launching
surface and over-expands. Later, it is forced back towards the axis
by magnetic hoop stress and produces a recollimation shock
\citep{BT05}. In all cases the shock is visible as a bright spot in
the synthetic maps.

We performed our parametric studies unbiased towards the existence of
a recollimation shock. All high ranking models have a recollimation
shock. Only after ranking our models, we favored for illustrative
purposes models with pronounced brightness increase at some point
along the axis. We therefore conclude, that the optical bright knot
HST-1 is a general feature of all models matching our selection
criteria and cannot be seen independently from them. In particular, we
suggest, that any physical model explaining the opening angle and
morphological structure will simultaneously also account for HST-1.

In addition our best fitting model is consistent with a number of
observational constraints such as the magnetic field in the knot
HST-1, and the jet-to-counterjet brightness ratio.

\acknowledgements 
The authors thank the anonymous referee for helping us
realize that our model also predicts an essentially one sided
jet as observed, 
Y.~Y.~Kovalev for providing helpful information on
the M87 inner jet structure prior to publication,
and S.~Jester for fruitful discussion of the manuscript. 
Part of this work
was supported by the European Community's Research Training Network
RTN ENIGMA under contract HPRN--CY--2002--00231 and by the European
Community's Marie Curie Actions - Human Resource and Mobility within
the JETSET (Jet Simulations, Experiments and Theory) network under
contract MRTN-CT-2004 005592. I.~A. acknowledges support by the
``Spanish Consejo Superior de Investigaciones Cient\'ificas'' through
an I3P contract and the ``Ministerio de Ciencia e Innovaci\'on'' and
the European fund for Regional Development through grant
AYA2007-67627-C03-03.

\bibliographystyle{apj}
\bibliography{references}

\end{document}